\documentclass[reprint,amsmath,amssymb,aps,prl,superscriptaddress]{revtex4-1}

\usepackage{graphicx}% Include figure files
\usepackage{dcolumn}% Align table columns on decimal point
\usepackage{bm}% bold math
\usepackage[dvipsnames]{xcolor}
\usepackage{soul}
\newcommand{\Dtm}{D^2}

\newcommand{\phiJ}{\phi_{\rm J}}
\newcommand{\phiG}{\phi_{\rm G}}

\renewcommand{\vec}[1]{\bm{#1}}

\newcommand{\appropto}{\mathrel{\vcenter{\offinterlineskip\halign{\hfil$##$\cr
    \propto\cr\noalign{\kern2pt}\sim\cr\noalign{\kern-2pt}}}}}

\begin{document}

\title{Slow coarsening in jammed athermal soft particle suspensions}

\author{R. N. Chacko}
 \affiliation{Department of Physics, Durham University, Science Laboratories,
  South Road, Durham DH1 3LE, UK}
%\author{R. Matas Navarro}
% \affiliation{Department of Physics, Durham University, Science Laboratories,
%  South Road, Durham DH1 3LE, UK}
\author{P. Sollich}
 \affiliation{University of G\"ottingen, Institute for Theoretical Physics, 37077 G\"ottingen, Germany}
 \affiliation{King's College London, Department of Mathematics, London WC2R 2LS, UK}
\author{S. M. Fielding}
 \affiliation{Department of Physics, Durham University, Science Laboratories,
  South Road, Durham DH1 3LE, UK}

\date{\today}

\begin{abstract}

We simulate a densely jammed, athermal assembly of repulsive soft particles immersed in a solvent. Starting from an initial condition corresponding to a quench from a high temperature,
we find non-trivial slow dynamics driven by a gradual release of stored elastic energy, with the root mean squared particle speed decaying as a power law in time with a fractional exponent. This decay is accompanied by the presence within the assembly of spatially localised and temporally intermittent `hot-spots' of non-affine deformation, connected by long-ranged swirls in the velocity field, reminiscent of the local plastic events and long-ranged elastic propagation that have been intensively studied in sheared amorphous materials. The pattern of hot-spots progressively coarsens, with the hot-spot size and separation slowly growing over time, and the associated  correlation length in particle speed increasing as a sublinear power law. Each individual spot however exists only transiently, within an overall picture of strongly intermittent dynamics.

\end{abstract}

\maketitle

The physics of amorphous yield stress materials such as colloids, microgels, onion surfactants, star polymers, emulsions and foams has been the focus of intense research in recent years~\cite{bonn2017yield}. In spite of similarities in their macroscopic behaviour there is  an important distinction within this class of systems, between thermal or athermal materials. Hard sphere (or soft, e.g.\ Lennard Jones) colloids that are small enough to experience Brownian motion undergo a thermal {\em glass} transition at a packing fraction $\phiG$. In contrast, soft particles (or droplets or bubbles) with radii $R\gtrsim 1\mu$m large enough  for Brownian motion to be negligible undergo an athermal {\em jamming} transition at a packing fraction $\phiJ>\phiG$~\cite{ikeda2012unified}.
The difference between the two types of system is apparent also in the order of magnitude of the yield stress above the transition: for thermal systems this is $k_{\rm B}T/R^3$ in line with the entropic origin of rigidity, whereas in the athermal case the yield stress is of the order of the singe particle modulus $G$. Intermediate particle sizes result in crossover behaviour between these limits~\cite{ikeda2013disentangling}.

The behaviour of {\em thermal} colloidal glasses in the absence of externally imposed flow has been studied in some detail~\cite{hunter2012physics}. They display slow relaxation dynamics characterised by the power law decay of one-time quantities,
and aging in two-time quantities such as the self-intermediate scattering function, mean squared particle displacement, or stress relaxation following a step strain~\cite{bandyopadhyay2010stress,kob2000aging,martinez2008slow}.
Local plastic particle rearrangements have been identified as a key relaxation mechanism ~\cite{lemaitre2014structural}. These appear as activated processes in the aging dynamics and are the analogues of particle rearrangement events seen in shear flow. In either case a local plastic event is followed by long ranged elastic stress propagation
~\cite{tanguy2006plastic,%
falk1998dynamics,%
chattoraj2013elastic,%
chikkadi2011long,%
picard2004elastic,%
schall2007structural,%
desmond2015measurement} as analysed for homogeneous materials by Eshelby~\cite{eshelby1957determination,picard2004elastic}. Under slow shear and at low temperatures this elastic propagation mechanism leads to plastic events organising into system-spanning avalanches that flicker in and out of existence across the sample~\cite{salerno2012avalanches,%
maloney2006amorphous,%
lemaitre2007plastic,%
bailey2007avalanche,%
dasgupta2012microscopic,%
lin2014scaling,%
puosi2014time,%
liu2016driving}.

In contrast, the physics of {\em athermal} soft suspensions in the absence of flow  has been assumed trivial to date. Lacking as these materials do any thermal agitation, it has been generally assumed that they should jam up exponentially quickly, on a timescale set by the interparticle solvent viscosity divided by the particle modulus, with each particle rapidly attaining a local energy minimum relative to its neighbours. Indeed, a theoretical search for slow relaxation modes in athermal foams proved fruitless~\cite{buzza1995linear}. Nonetheless, experimentally measured viscoelastic spectra in foams, which are athermal on account of their typical large bubble sizes ($R\approx 30\mu$m in~\cite{khan1988foam}), do reveal characteristically flat loss moduli $G''(\omega)$ down to the lowest accessible frequencies, indicating the presence of very slow relaxation modes~\cite{khan1988foam,footnote1}.
    
The contribution of this Letter is to show by detailed computer simulations that athermal suspensions of soft particles {\em do} show non-trivial dynamics without externally imposed flow. Following sample preparation at some initial time $t=0$ in a state with packing fraction $\phi>\phiJ$, we demonstrate a scenario of slow dynamics driven by a progressive decline of the stored elastic energy per particle, $V(t)$, towards a final value $V_\infty$. The associated rate of change $-\dot{V}(t)$ decays as a power law $t^{-\alpha}$ with a fractional exponent $\alpha$, as does the root mean squared particle speed, $v_{\rm rms}(t)~\sim t^{-\beta}$. Such a scenario suggests that the dynamics within the assembly become correlated on progressively larger lengthscales as time proceeds. Indeed, we identify the presence of temporally intermittent and spatially localised `hot-spots' of non-affine deformation, connected by long-ranged swirling velocity vortices, in a scenario reminisent of the elastoplastic Eshelby dynamics of sheared amorphous materials~\cite{tanguy2006plastic,falk1998dynamics,chattoraj2013elastic,chikkadi2011long,picard2004elastic,schall2007structural,desmond2015measurement}. The hot-spot pattern slowly coarsens over time, with the hot-spots gradually becoming larger and further apart, although each individual spot itself exists only transiently. The associated  correlation length in particle speed grows as a sublinear power law, $l^*\sim t^{\gamma}$.     

We simulate an athermal assembly of  repulsive soft particles immersed in a solvent at high packing fraction $\phi>\phi_{\rm J}$, assuming a pairwise repulsive harmonic interaction potential, $V_{ij}(r)=GR^3 (1-r/D_{ij})^2\Theta(D_{ij}-r)$, where 
$\Theta(x)$ is the Heaviside
function. We take a bidisperse mixture of $N=10^6$ particles with radii $R$ and $1.4R$ in equal number, with $D_{ij}=R_i+R_j$ for any particle pair $ij$.  Neglecting explicit hydrodynamic interactions, we assume that the potential forces are balanced simply by drag against the solvent, with the position of the $i$th particle obeying:
\begin{equation}
\label{eqn:dynamics}
    \frac{d\vec{r}_i}{dt}
    =-\frac{1}{\zeta}\sum_{j\neq i}\frac{\partial V(|\vec{r}_i-\vec{r}_j|)}{\partial \vec{r}_i},
\end{equation}
where $\zeta$ is the drag coefficient.
This corresponds to gradient descent, which for finite $N$ will terminate in a local energy minimum.
(It should be noted that the form of drag in Eqn.~(\ref{eqn:dynamics}), although used widely in the literature, violates Galilean invariance. However, we have checked that replacing it by the properly invariant -- although still approximate -- form used in dissipative particle dynamics~\cite{groot1997dissipative} does not change the physical scenario that we report below.) Such a simulation is intended to describe, for example, a dense emulsion, foam or microgel comprising droplets, bubbles or particles of typical radius $R\gtrsim 1\mu$m, in the athermal regime.
Hereafter, we shall refer to bubbles and droplets also simply as particles.
\begin{figure}[!t]
\hbox{\hspace{-0.7em} \includegraphics[width=0.49\textwidth]{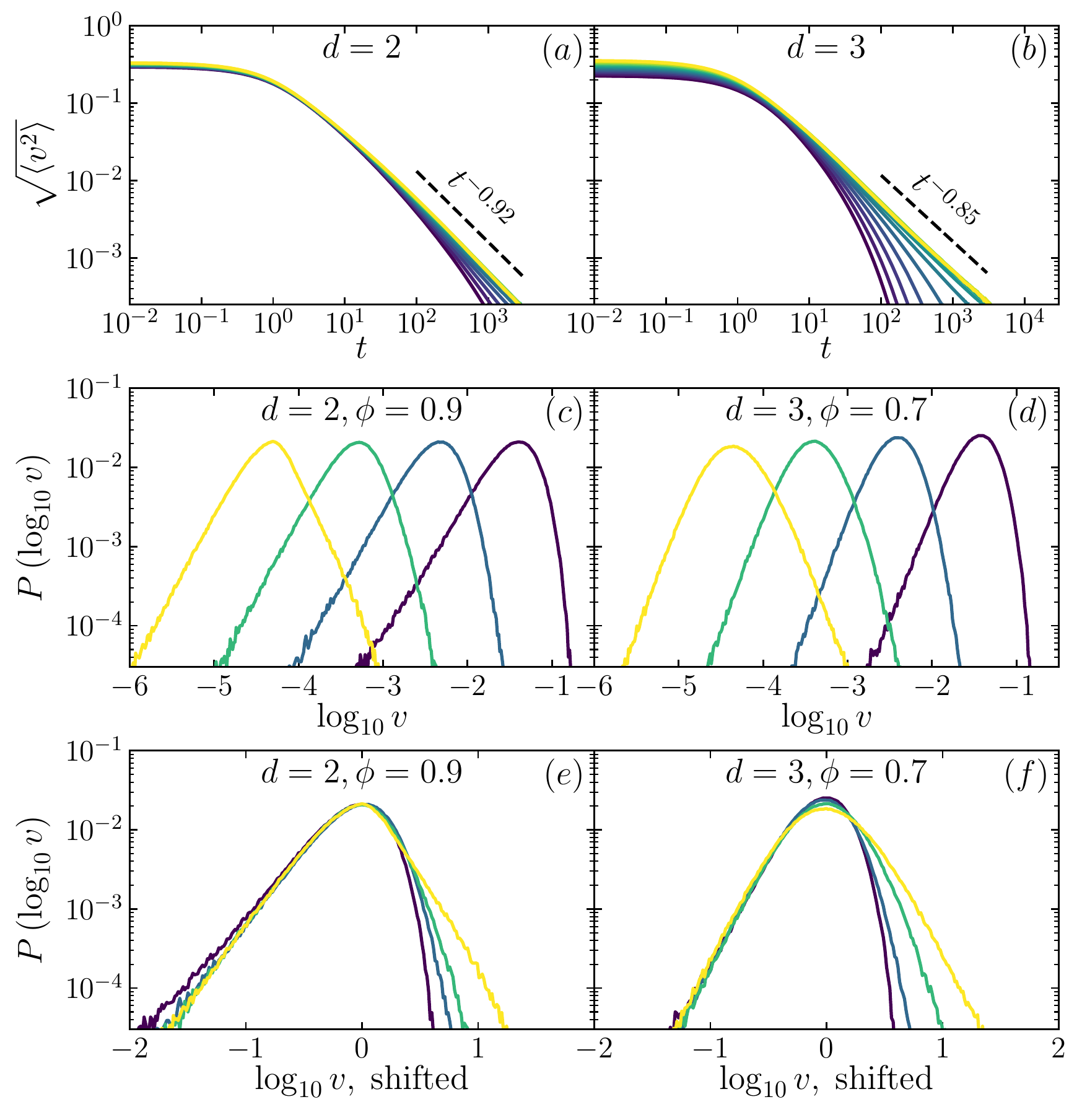}}
\caption{Decay of the root mean squared particle speed in (a) $d=2$, for $\phi=0.78, 0.80 \ldots 0.98, 1.00$ (curves upwards)
and (b) $d=3$, for $\phi=0.40, 0.45, \ldots 0.95, 1.00$ (curves upwards).
Distributions of log particle speed for $d=2,\phi=0.90$ (c) and $d=3,\phi=0.70$ (d), both for times $t=10^1,10^2,10^3$ and $10^4$ (curves leftward).
(e) and (f): Distributions shown in (c) and (d) respectively, shifted horizontally to highlight their broadening with increasing time.} 
\label{fig1}
\end{figure}

We integrate Eqn.~(\ref{eqn:dynamics}) using an explicit Euler algorithm, with an adaptive timestep scaled by the inverse maximum particle speed, $\Delta t(t)=a/v_{\rm max}(t)$. All results are converged to the limit $a\to 0$. We choose units of length in which the smaller particle radius $R=1$, of time in which the drag coefficient $\zeta=1$, and of mass in which the particle modulus $G=1$. The physical parameters that remain to be explored are then the packing fraction of the assembly, $\phi$, and the spatial dimensionality, $d$. We mostly present results below for $d=2$, with supplementary data in~\cite{SuppMat} demonstrating that the same physical scenario also holds for $d=3$.

Two different initial conditions are considered. In the first, the particles are placed in the simulation box at time $t=0$ with uniformly distributed positions. Physically, this corresponds to a sudden quench to temperature $T=0$ from a previously infinite temperature. In the second, the assembly is initially equilibrated at a volume fraction below jamming, $\phi=0.5<\phiJ$, and nonzero temperature, $T=0.01$, implemented by augmenting Eqn.~(\ref{eqn:dynamics}) with a thermal noise term. The particles are then suddenly expanded in situ at time $t=0$ to achieve the desired $\phi>\phiJ$, and temperature is set to $T=0$. This is intended to model the rapid swelling of core-shell particles to reach a jammed state.
%~\cite{}
Beyond an initial transient up to a typical time $t=10^2$, we find the same dynamical scenario including non-trivial power-law decays for both of these initial conditions, as shown in the supplemental material~\cite{SuppMat}. Accordingly, results will be presented below only for the fully random initial condition.

\begin{figure}[!t]
\hbox{\hspace{-0.6em} \includegraphics[width=0.49\textwidth]{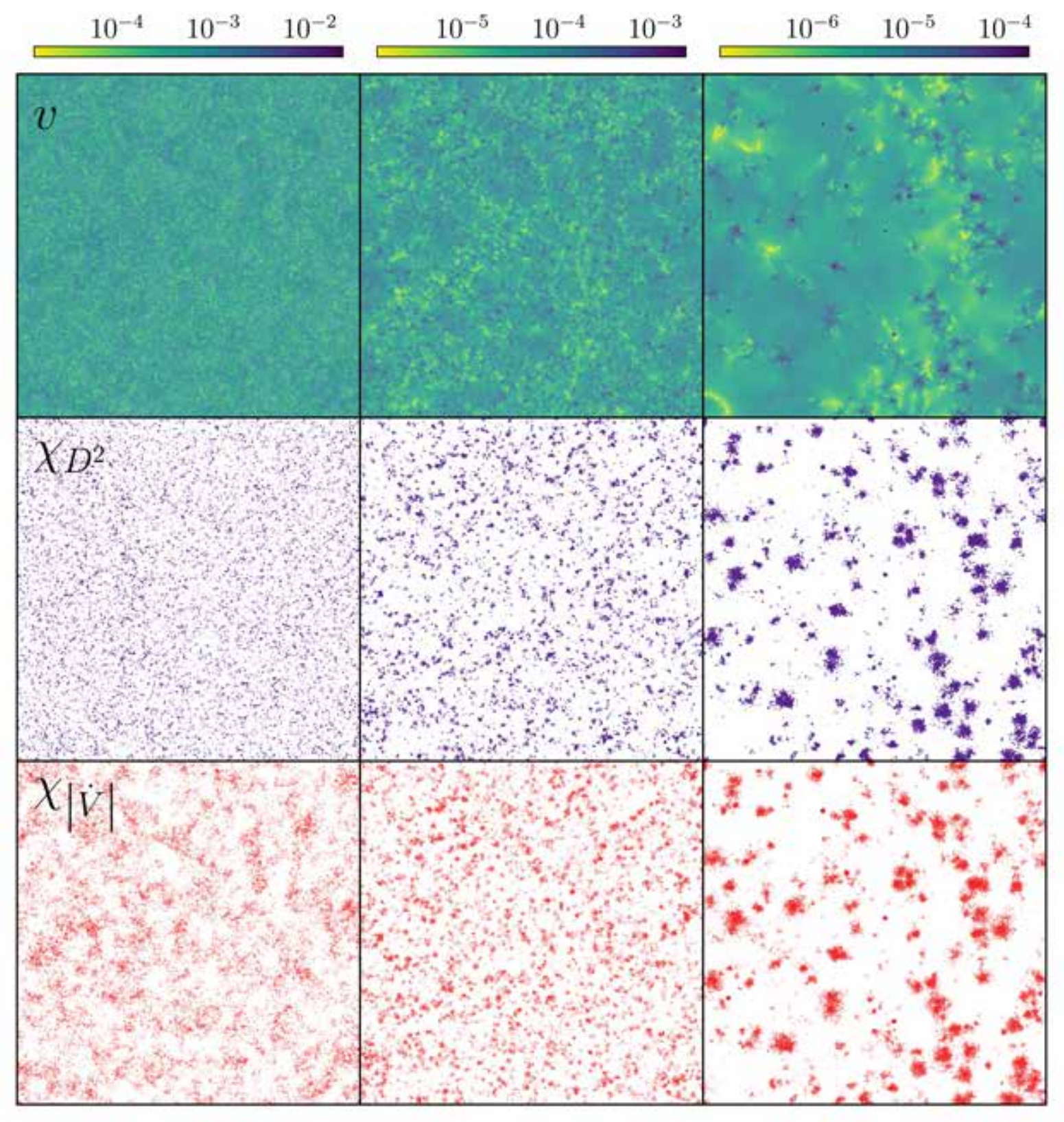}}
\caption{Colour maps in columns rightwards for times $t=10^3, 10^4$ and $10^5$ of (top) particle speeds, (centre) thresholded non-affine deformation rate $\chi_{D^2}$ and (bottom) thresholded magnitude of rate of change of particle energy $\chi_{|\dot V|}$.
The colour scale for $v$ is centred around the most likely value of $\log_{10}v$.
Thresholded plots show particles in dark if they are within the top 10\% as ranked by the quantity in question, white otherwise. }
\label{fig2}
\end{figure}

We start by showing in Fig.~\ref{fig1} the decay of the root mean squared particle speed for a range of packing fractions $\phi$ from just below to far above $\phiJ$. We do so for both $d=2$ (Fig.~\ref{fig1}a), where $\phiJ=0.84$ and for $d=3$ (Fig.~\ref{fig1}b), where $\phiJ=0.64$. After an initial transient persisting to a time $t=O(1)$, the decay enters a power law regime. For packing fractions below jamming, $\phi<\phiJ$, this power law is eventually cut off by a final exponential decay. Above jamming, $\phi>\phiJ$, the power law $t^{-\alpha}$ persists to a time that increases without bound as the system size $N$ increases. (For any finite $N$, it finally ends in noise~\cite{SuppMat}. We show in Fig.~\ref{fig1} only $N$-independent results.) We find an exponent $\alpha=0.92$ in $d=2$ and $\alpha=0.85$ in $d=3$. In both cases the typical speeds therefore decay significantly more slowly than one would expect for trivial non-glassy decay towards a crystalline lattice, where a simple analytical calculation yields $\alpha=1+d/4$.
The associated log-speed distribution gradually shifts to lower speeds overall, but with a tail for higher than average speeds that also broadens slightly, Fig.~\ref{fig1}c--d.

The observation of a steadily slowing power law decay suggests that spatial correlations develop within the assembly on progressively larger lengthscales as time proceeds. To investigate this, we show in Fig.~\ref{fig2} snapshot colour maps of the particle speed $v_i$, the temporal rate of change of particle energy $\dot{V}_i$ (where $V_i=\sum_{j\neq i} V(|\bm{r}_i-\bm{r}_j|)/2$ is the potential energy contribution of particle $i$), and the non-affine part of the local deformation rate. The latter is defined for each particle $i$ as 
\begin{equation} \label{eqn:nonaffine}
\Dtm := \min\limits_{\bm{K}}  \sum\limits_{j \neq i} \left| \left| (\bm{v}_i-\bm{v}_j) - \bm{K}(\bm{r}_i-\bm{r}_j)\right| \right|^2,
\end{equation}
where the sum runs over neighbours within a maximum distance of $3R$.
This is an instantaneous (short time interval) version of $D_\mathrm{min}^2$ from~\cite{falk1998dynamics}.
For ease of visualisation we define thresholded versions of $\Dtm$ and $|\dot{V}|$, denoted $\chi_{D^2}$ and $\chi_{|\dot{V}|}$ respectively, equal for each quantity to $1$ for particles in the top $10\%$ of the distribution of  values of that quantity, and $0$ otherwise.
At any fixed time, localised hot-spots of high particle speed, rapidly changing particle energy, and high non-affine deformation rate are evident. With increasing time in rows downwards in Fig.~\ref{fig2}, the patterns coarsen as the hot-spots become progressively larger and further apart. Each spot itself exists only transiently, however, with a dynamics that is highly intermittent in nature. We shall return to expand on this point below.

\begin{figure}[!t]
\centering
\includegraphics[width=0.49\textwidth]{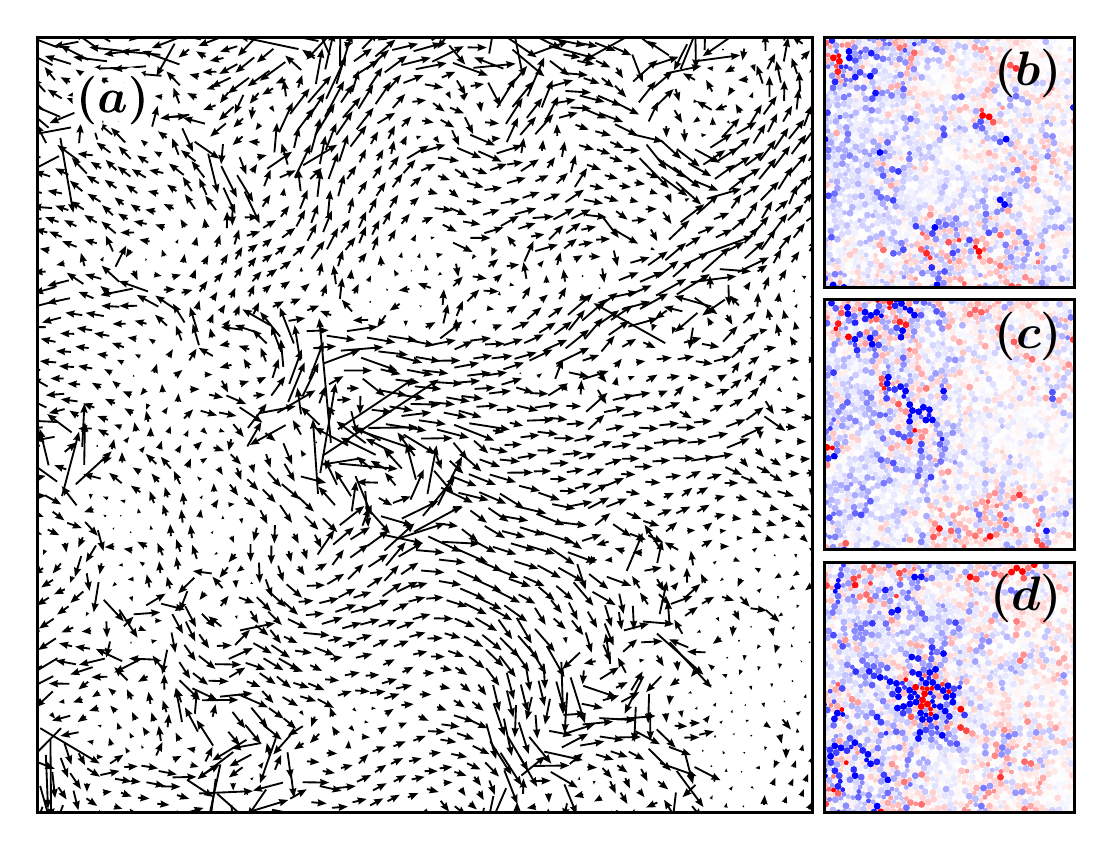}
\caption{Zoom of a $100\times 100$ region for a $d=2$ system of $N=10^6$ particles at packing fraction $\phi=0.9$.
(a) Displacement vectors (scaled up by a factor 25) over the time interval $1.8\times 10^3$ to $2.2 \times 10^3$.
(b--d) Dynamical intermittency, seen in the rate of change of particle energy $\tilde{\dot{V}}$ at times $t=1148.75$, $1518.29$ and $1984.45$ respectively.
Colour scale linearly interpolates from $\tilde{\dot{V}}= -20$ (blue) to $\tilde{\dot{V}}=20$ (red).}
\label{fig:swirls}
\end{figure}

This scenario of spatially localised hot-spots is strongly reminiscent of the local plastic events (sometimes called shear transformation zones) that are widely discussed in the context of sheared amorphous materials~\cite{tanguy2006plastic,falk1998dynamics,chattoraj2013elastic,chikkadi2011long,picard2004elastic,schall2007structural,desmond2015measurement}, and that also arise via thermal activation in unsheared colloids~\cite{lemaitre2014structural}. Each such zone has been shown to create around itself a long-ranged Eshelby perturbation according to the Green's propagator of linear elasticity,
giving rise to long-ranged vortex-like structures in the velocity field. Strikingly similar behaviour is seen in our simulations: a typical map of displacement (integrated over a short time interval) is displayed in Fig.~\ref{fig:swirls} and clearly resembles that of  Fig.~3 of Ref.~\cite{maloney2006amorphous} for a sheared  amorphous material, with the overall imposed shear subtracted out.

\begin{figure}[!t]
\includegraphics[width=0.49\textwidth]{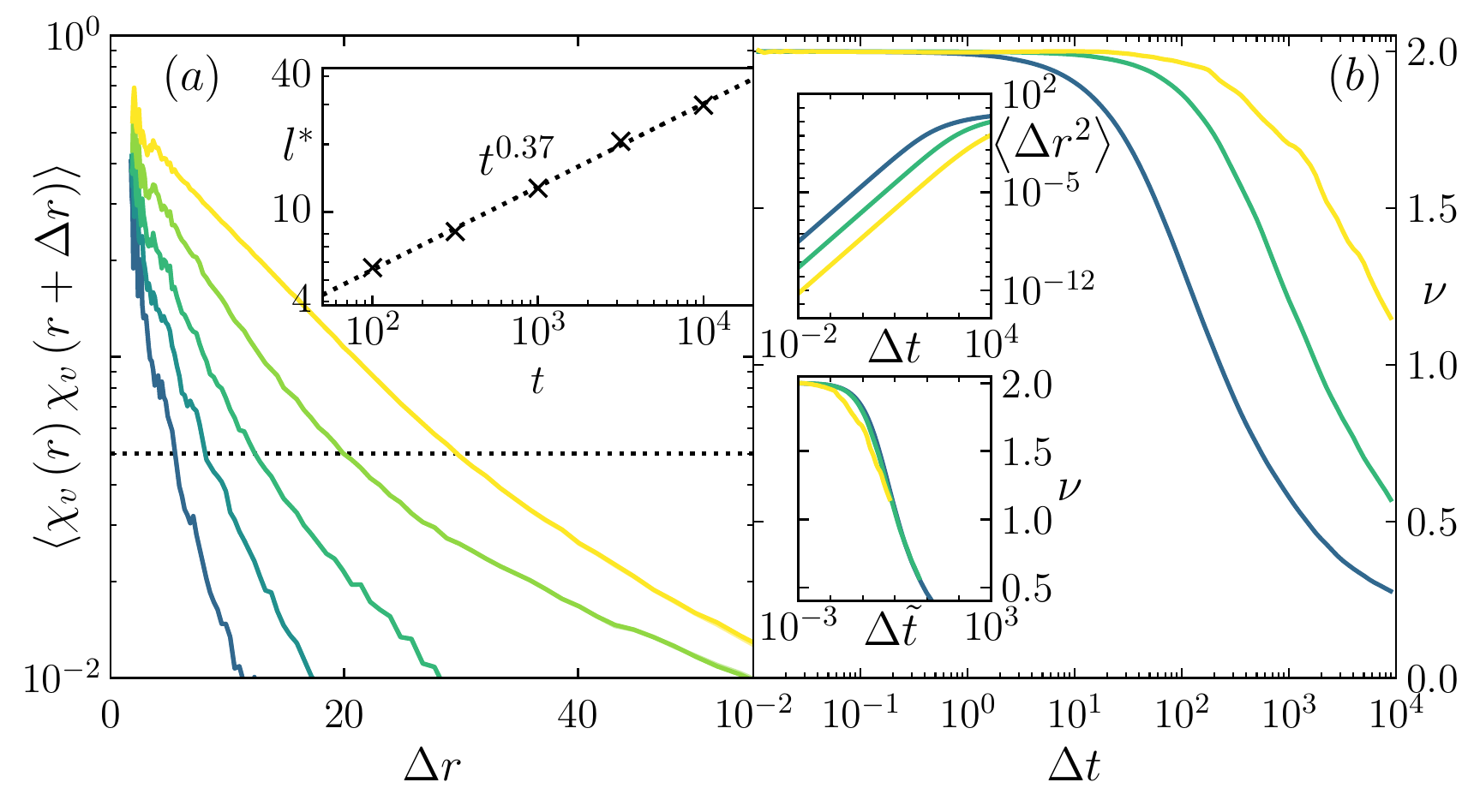}
\caption{(a) Normalised spatial correlation function of the thresholded particle speed $\chi_v$, at times $t=10^2,10^3$ and $10^4$ (curves left to right).
Inset: lengthscale $\Delta r=l^*$ at which the correlator falls below $0.05$, shown as a function of time. Dashed line is a fit to the power law $t^{0.37}$.
(b) Aging behaviour of the mean-squared displacement $\left< \Delta r^2\right>$ of particles between times $t$ and $t+\Delta t$, shown as a function of the time interval $\Delta t$ for several $t$. The raw data are shown in the top inset, with curves rightwards corresponding to $t=10^2, 10^3, 10^4$. Main panel: 
exponent $\nu
\equiv \partial \ln \left< \Delta r^2\right> / \partial \ln \Delta t$ against $\Delta t$.
Bottom inset shows $\nu
$ against $\Delta t \sqrt{\left< v^2 \right>}$, collapsing the ballistic parts of the curves. All data are for $d=2$, $\phi=0.9$.
} 
\label{fig4}
\end{figure}

To investigate further the structure of these swirling vortices, we show in Fig.~\ref{fig4}a the spatial correlation function of the particle speed,
again thresholded on the top $10\%$ to give  $\chi_{v}$, for values of time increasing in curves rightwards.
Its characteristic decay length increases over time according to a power law $l^*\sim t^\gamma$ where we estimate the exponent as $\gamma=0.37$ in $d=2$,
Fig.~\ref{fig4}a (inset); from variation of the thresholding procedure and comparison with unthresholded velocity correlations we estimate the exponent uncertainty to be about $0.04$ (see supplemental material~\cite{SuppMat}).
In Fig.~\ref{fig4}b we demonstrate that this correlation length increase is associated with a characteristic aging behaviour in the mean squared displacement
$\langle \Delta r^2\rangle$ of particles.
Specifically, the exponent $\nu$ governing the increase in $\langle \Delta r^2\rangle$ with time difference $\Delta t$ crosses over from
the ballistic value $\nu=2$ to subdiffusive growth $\nu<1$ on a timescale that grows with age $t$ (Fig.~\ref{fig4}b, lower inset).

\begin{figure}[!t]
\includegraphics[width=0.49\textwidth]{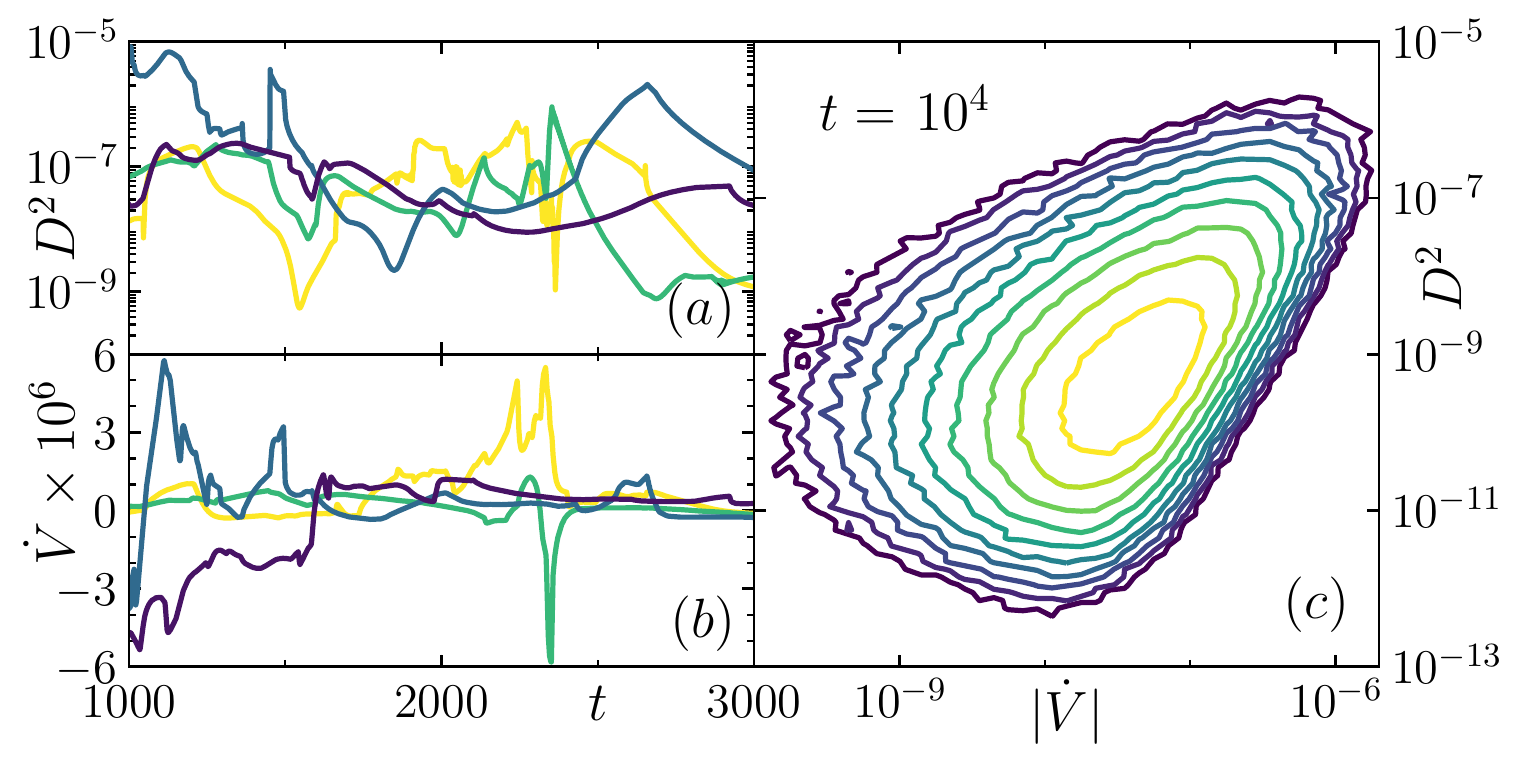} % Scale factor needed to minimise blurriness
\caption{(a) Time signal of the non-affine deformation rate, defined in Eqn.~(\ref{eqn:nonaffine}) for
four particles (chosen out of one hundred randomly selected particles such that their $\dot V$ dynamics is of a similar scale).
(b) Corresponding signal of the rate of change of particle energy for the same four particles.
(c) Correlation of the non-affine deformation rate with the absolute rate of change of particle energy at time $t=10^4$.
Contours show level sets for the number of particles, ${N=10^{1.5},10^{1.7},\cdots,10^{3.5}}$,
within bins of width $0.2$ and height $0.4$
in $\left( \log \left| \dot V \right|, \log \left| \Dtm \right| \right)$-space.
Data shown for packing fraction $\phi=0.9$ in two dimensions, $d=2$.}
\label{fig5}
\end{figure}

So far, we have shown that the pattern of hot-spots slowly coarsens over time, with increasing hot-spot size and separation, and a growing correlation length $l^*(t)$ in the particle speed. Alongside this overall picture, we now further demonstrate the dynamics to be strongly intermittent, with each individual hot-spot itself existing only transiently. This can already be seen in the colour maps of the rate of change of particle energy in Fig.~\ref{fig:swirls}b-d. We further show in Fig.~\ref{fig5}a,b the temporal signals of non-affine deformation rate, and the rate of change of particle energies, for four individual particles. Each particle clearly experiences intervals of relative quiescence punctuated by bursts of high local activity. The correlation between local non-affinity and rate of change of particle energy that is visually apparent in the time signals of Fig.~\ref{fig5}a,b is demonstrated quantitatively in the contour plot of Fig.~\ref{fig5}c.

To summarise, we have demonstrated a scenario of non-trivial slow dynamics in a dense athermal suspension of purely repulsive soft particles,  with the mean squared particle speed decaying over time as a sublinear power law. This contrasts with the naive intuition that any such system should quickly jam up as each particle rapidly attains a local energy minimum relative to its neighbours, and indicates that collective effects become increasingly dominant over time, on progressively larger lengthscales. Consistent with this expectation, we have demonstrated the existence within the assembly of hot-spots of locally non-affine deformation, linked by long ranged swirls in the velocity field, in a scenario reminiscent of the local plastic events and elastic stress propagation that has been intensively studied in sheared amorphous materials. The pattern of hot spots slowly coarsens over time, with the associated velocity correlation length increasing as a sublinear power law.

There is clearly significant scope for further investigation of the non-trivial aging dynamics we have identified in athermal systems. The structure and growth of the hot-spots would be interesting to study, for example. As the system descends in its potential energy landscape, less and less energy is available to `activate' plastic rearrangements over local energy barriers. One would then conjecture that these plastic rearrangements, which take place within the hot-spots, become increasingly collective in order to reduce the height of the corresponding energy barriers.
The spectrum and properties of instantaneous normal modes during the aging process will also be interesting to study, in particular in relation to existing results for statically jammed or driven athermal systems~\cite{OHeLanLiuNag01,LiuNag10}
Finally, the consequences of athermal aging for the rheology (mechanical behaviour) of these materials will be fascinating to explore. This will provide non-trivial benchmarks on which to test and refine existing mesoscopic modelling approaches~\cite{HebLeq98,SolOliBre17,SolLeqHebCat97,FalLan11} 
in the relevant athermal regime. It will also help us to understand the qualitative differences and commonalities with aging in thermal, largely entropy-driven glasses such as colloidal hard spheres, where mode-coupling theory has been shown to be a useful modelling paradigm~\cite{VanPus91,BraVoiFucLarCat09,Kob97}.
 
 {\it Acknowledgements ---} The authors thank Mike Cates for useful discussions.
The research leading to these results has received funding from SOFI CDT, Durham University, from the EPSRC (grant ref.~EP/L015536/1) and from the European Research Council under the European Union's Seventh Framework Programme (FP7/20072013) / ERC grant agreement number 279365.

\bibliographystyle{apsrev4-1}

\end{document}